\documentclass{article}
\usepackage{graphicx}

\usepackage{epstopdf}

\begin{document}
\begin{titlepage}

\begin{center}
{\large \textbf{Oscillating Entropy}}

\vspace{2\baselineskip}
{\sffamily Enrique Canessa\footnote{e-mail: canessae@ictp.it}}

{\it The Abdus Salam International Centre for Theoretical Physics, Trieste, Italy}

\vspace{2\baselineskip}
\end{center}

\begin{abstract}
The log-periodic equation for the entropy
$S = - (k/a) \sum_{i=1}^{N} p_{i} \sin(a \ln p_{i})$, based on the forgotten Sharma-Taneja 
entropy measure, is studied for the first time with $N$ the total number of system states
and $p_{i}$ the associated probabilities.
 It is argued that an oscillatory regime for $S$ could in principle be
understood in terms of a linear time-dependent behavior for the associated probabilities
in analogy with a spring system of frequency $w$ gaining momentum from the surroundings. The
physical meaning for the production of entropy given by the parameter $a$ relates
the angle $\omega t$. We discuss its properties, and make a connection with the 
non-extensive Tsallis, R\'{e}nyi, Boltzmann-Gibbs and Shannon entropies as special limiting cases
for systems with constant mass. Our non-trivial form of entropy displays peculiar concavity and 
addition properties. Log-periodicity in $S$, concavity lost, and non-additivity are manifested by increasing 
the value of the coefficient $a$, which sets the variations with respect to the behavior of the monotonic
Gibbs entropy function. We have considered systems with linearly increasing mass and have explained 
how "oscillating entropy" in such systems may appear.

\end{abstract}

\vspace{2\baselineskip}

\textbf{PACS numbers:} 89.70.Cf, 89.75.-k, 05.20.-y, 05.70.-a

\end{titlepage}
\newpage


\section{Introduction}

Almost two decades ago, Esteban and Morales \cite{Est95} reported a summary of entropy 
statistics with the purpose of comparing a large amount of entropy measures within a simple probability 
functional --where particular entropy expressions are obtained as limiting cases by adjusting two 
parameters $(r,s)$. Each having some advantages over others, either in
their simplicity or due to opening a new door for physical interpretation.
Most of such listed entropy functions, {\it e.g.}, the Gibbs-Shannon and R\'{e}nyi formula \cite{Ren88}, 
continue to be extensively discussed in the literature. Included in their study \cite{Est95} was also the 
alternative Sharma-Mittal entropy for an arbitrary exponential family of distributions \cite{Nie12}, which generalize 
the popular Tsallis-class of functions in some limiting case, $r=1$ \cite{Akt07}. Curiously, there is another particular 
entropy function listed in \cite{Est95}, proposed few years earlier by Sharma and Taneja \cite{Sha77}, that has been 
completely ignored in the literature.  This is given by
\begin{equation}\label{eq:logperiod_st}
S(x) = - (k/sin (s)) x^{r} \sin(s \ln x) \;\;\; .
\end{equation}
This expression has received little attention by physicists and it consists in the introduction of an oscillatory 
(sine) function driven by the $r$ and $s$ parameters.

Generalizations of the famous Gibbs probabilistic description of entropy given by
$S^{G} = - k \sum_{i=1}^{N} p_{i} \ln p_{i}$,
usually incorporate extra parameters to describe a wide spectrum of complex phenomena.  Examples
are the Tsallis entropy function based on multifractal scaling concepts $p_{i}^{q}$
\cite{Tsa88}, and the $q$-tagged R\'{e}nyi formula \cite{Ren88}. The modeling of the $q$ parameters
in these generalized entropy forms --also appearing in the most recent deformed entropy formulas
incorporating zero mutual information between finite reservoir effects \cite{Bir13}, still deserves further
study \cite{Bir11,Tsa12}.  As recently pointed out in \cite{Tsa13} (see also \cite{Ola01,Can04,Zhe13}), entropies generalizing that of 
Boltzmann-Gibbs become necessary in order to recover thermodynamic extensivity for non-standard systems.

Motivated by the need of a model to understand the universal nature of $S$ from a microscopic point of view \cite{Bir11},
in this work we shall study the log-periodic entropy equation of Sharma and Taneja \cite{Sha77} for the first
time so far as we know. 
We shall consider the log-periodic entropy function of eq.(\ref{eq:logperiod_st}), with $r=1$, to show 
that this form of entropy displays peculiar concavity and addition properties.
We also show that variations with respect to the behavior of the monotonic Gibbs entropy function
appears as log-periodicity in $S$, concavity lost, and non-additivity.
We argue that an oscillatory regime for $S$ is possible which can in principle be 
understood in terms of a linear time-dependent behavior for the associated probabilities
in analogy with a spring system gaining momentum from its surroundings as introduced by 
the author \cite{Can09}.  This approach may help to figure out the 
nature of entropy from a new perspective for variations in the system mass.


\section{Sharma-Taneja Entropy ($r=1$) and Physics Interpretation}

The restoring force $F$ that a spring exerts on a fluctuating mass 
$m_{t}$ resting on a frictionless surface and attached to one end of 
the spring is defined by Hooke's law.  For small changes in lenght this is given by
\begin{equation}\label{eq:force}
F = {d (m_{t}v) \over dt} = 
  \left({dm_{t} \over dx}\right) \left({dx \over dt}\right)^{2} + 
     m_{t} \left({dv \over dt}\right) \equiv  -kx \;\;\; ,
\end{equation}
where $x$ is the displacement from the spring equilibrium 
position $x_{0}$ at time $t_{0}$, $v = dx/dt$ is the system 
velocity and $k$ an spring constant.  

For $t \ge t_{0} \ne 0$, and considering a linear fluctuating mass with time 
$m_{t} = (t/t_{0}) m_{0}$ and the spring constant inversely 
proportional to $t$, $k_{t} = (t_{0}/t) k_{0}$,
then the general solution of this equation has the form (see \cite{Can09,Oze10})
\begin{equation}\label{eq:logp}
x(t) = x_0 \; sin\left[ \theta\; ln \left({t \over t_0}\right) \right] +  
 x_1 \; cos\left[ \theta\; ln \left({t \over t_0}\right) \right] \;\;\; ,
\end{equation}
with $x_{0,1}$ constants.  

Hence it follows that classical and quantum oscillator systems
lead to temporal log-periodic oscillations as a consequence of adding mass linearly with time.
If we assume therefore that the evolution for a microscopic system gains momentum from its surroundings and it satisfies a linear 
time-dependent behavior for its associated state probabilities $p$, then the entropy function of our system
may display log-periodicity when considered as expectation values $E[\;f(p(t))\;]$.

Within this approach, let us consider for the entropy of complex systems
\begin{equation}\label{eq:logperiod}
S = - (k/a) \sum_{i=1}^{N} p_{i} \sin(a \ln p_{i}) \;\;\; ,
\end{equation}
where the sum runs over all possible $N$ distinct states of the system, and $p_{i}$ is the probability that the
microsystem could be found in the $i$-th state.
This expression corresponds to the limiting Sharma and Taneja entropy measure in eq.(\ref{eq:logperiod_st}) with $r=1$.
Since $0 < p_{i} \le 1$ and $\sum_{i=1}^{N} p_{i} \equiv 1$, then $S \ge 0$, $\forall |a| < \pi$.

In this case, the non-additivity property is described by
\begin{eqnarray}\label{eq:nonadd}
S^{A+B} & =  & - (k/a) \sum_{i=1}^{N_{A}}\sum_{j=1}^{N_{B}} p_{i}^{A}\cdot p_{j}^{B} \sin(a \ln (p_{i}^{A}\cdot p_{j}^{B}) ) \;\;\; ,  \nonumber \\
    & =  &  S^{A} \cdot \sum_{j=1}^{N_{B}} p_{j}^{B} \cos(a \ln (p_{j}^{B}) ) + S^{B}\cdot \sum_{i=1}^{W_{A}} p_{i}^{A} \cos(a \ln (p_{i}^{A}) ) \;\;\; ,  \nonumber \\
    & =  &  S^{A} + S^{B} - 2 \left\{ S^{A} \cdot \sum_{j=1}^{N_{B}} p_{j}^{B} \sin^{2}((a/2) \ln (p_{j}^{B}) ) \right. \nonumber \\
    &     & \left.  \hspace{2.8cm} + \; S^{B}\cdot \sum_{i=1}^{N_{A}} p_{i}^{A} \sin^{2}((a/2) \ln (p_{i}^{A}) )  \right\}  \;\;\; ,
\end{eqnarray}
where we have used the two trigonometric relations $\sin(\alpha + \beta) =\sin \alpha \cos \beta + \sin \beta \cos \alpha$ and $\cos \alpha = 1 - 2 \sin^{2} (\alpha/2)$.

The maximum value of each term in the generalized $S$ function of eq.(\ref{eq:logperiod})
satisfies $\tan(a \ln p_{m}) = -a$, which for $a \rightarrow 0$, and considering the maximum value found for the Gibbs entropy $S_{m}/k = p_{m} = 1/e$, it
results in the Taylor expansion $\tan (a) \approx a$. In this limit, from eq.(\ref{eq:logperiod}) we then have
\begin{equation}\label{eq;agibbs}
 \lim_{a \to \pm 0} S_{m}/S_{m}^{G} \approx \sin (a) / a \;\;\; .
\end{equation}
This positive relation provides insight into the meaning of the scalar coefficient $|a|<\pi$ appearing in our master entropy equation. Its value sets the amplitude of
the postulated log-periodicity for $S$ and variations with respect to the behavior of the monotonic Gibbs entropy function $S^{G}$. By l'Hopital's theorem we have
$S_{m} \approx S_{m}^{G}$ for $a \rightarrow 0$.
In this limit the Gibbs entropy of a system in thermal equilibrium with a thermal reservoir is recovered.
As shown in \cite{Can09}, the physical meaning for our parameter $a$ can be then understood in terms of the {\it constant} angle
\begin{equation}\label{eq:omega}
a \rightarrow \theta \equiv \omega_t \cdot t = \sqrt{{k_0 \over m_0}} \; t_0 
             = \omega_0 \cdot t_0 \equiv \theta_0 \;\;\; .
\end{equation}


\subsection{Continuous Log-Periodic Entropy}

Let us consider the concept of continuous or differential entropy.
The corresponding formula is defined using our form of the entropy as an expectation such that
\begin{equation}\label{eq:logperiodcont}
H = - (k/a) E[\sin(a \ln p(x))] = - (k/a) \int_{-\infty}^{\infty} p(x) \sin(a \ln p(x)) dx  \;\;\; .
\end{equation}
As with its discrete analog we need to take probability density functions $p(x)$, which in this case
can be greater than one.
With a standard normal distribution $p(x) = (1/\sqrt{2\pi\sigma^{2}})\exp(-(x-\mu)^{2}/2\sigma^{2})$, this differential 
entropy after some algebra has the following maximazed values
\begin{eqnarray}\label{eq:differential}
 (a/k) \sqrt{2\pi\sigma^{2}} H & =  & - \int_{-\infty}^{\infty} \exp(-(x-\mu)^{2}/2\sigma^{2}) 
    \sin(a [ -(1/2)\ln (2\pi\sigma^{2}) -  (x-\mu)^{2}/2\sigma^{2} ] ) \ dx  \;\;\; ,  \nonumber \\
    & =  &  \sin( (a/2)\ln (2\pi\sigma^{2}) ) \int_{-\infty}^{\infty}  \exp(-(x-\mu)^{2}/2\sigma^{2}) \cos( a (x-\mu)^{2}/2\sigma^{2} ) \nonumber \\
    &     &  + \; 
  \cos( (a/2)\ln (2\pi\sigma^{2}) ) \int_{-\infty}^{\infty}  \exp(-(x-\mu)^{2}/2\sigma^{2}) \sin( a (x-\mu)^{2}/2\sigma^{2} )   \;\;\; .
\end{eqnarray}
By a change of variables $x-\mu = \sqrt{t-\mu}$, with $\mu < t$, and expanding $\sin$ and $\cos$ in series, it 
can be shown that our continous entropy in sums of the Gammma function reaches the value
\begin{eqnarray}
 (\sqrt{\pi}/k) H & =  & 
     \sin( a\ln \sqrt{2\pi\sigma^{2}} ) \sum_{n=0}^{\infty}(-1)^{n} a^{2n-1} \Gamma (2n + 1/2)/(2n)!  \nonumber \\
    &     &  \hspace{0.5cm} + \; 
     \cos( a\ln \sqrt{2\pi\sigma^{2}} ) \sum_{n=0}^{\infty}(-1)^{n} a^{2n} \Gamma (2n + 3/2)/(2n+1)!   \;\;\; .
\end{eqnarray}
In first approximation ({\it i.e.}, $n=0$), this gives
\begin{equation}\label{eq:differential0}
H/k \approx (1/a) \sin( a\ln \sqrt{2\pi\sigma^{2}} ) + (1/2) \cos( a\ln \sqrt{2\pi\sigma^{2}} ) \;\;\; .
\end{equation}

As shown above, mathematical manipulations with a log-periodic function are not trivial. 
Log-periodicity of the form $sin(\ln x)$
appears in a number of other non-linear phenomena for an enhancement of predictions in different
contexts \cite{Sor02}, including econophysics \cite{Bar05,Dro03,Can09}, earthquake prediction
\cite{Sor96a,Can00a}, cooperative information \cite{Sor02a}. A connection between energy-spectrum
self-similarity and specific heat log-periodicity is discussed in \cite{Val98,Soa03}. We analyze
here the non-monotonic behavior of a log-periodic entropy function and its relation
with the non-extensive Tsallis, R\'{e}nyi, Boltzmann-Gibbs and Shannon entropies.


\section{Results and Discussion}

\begin{figure}[!t]
\centering
\includegraphics[width=17cm]{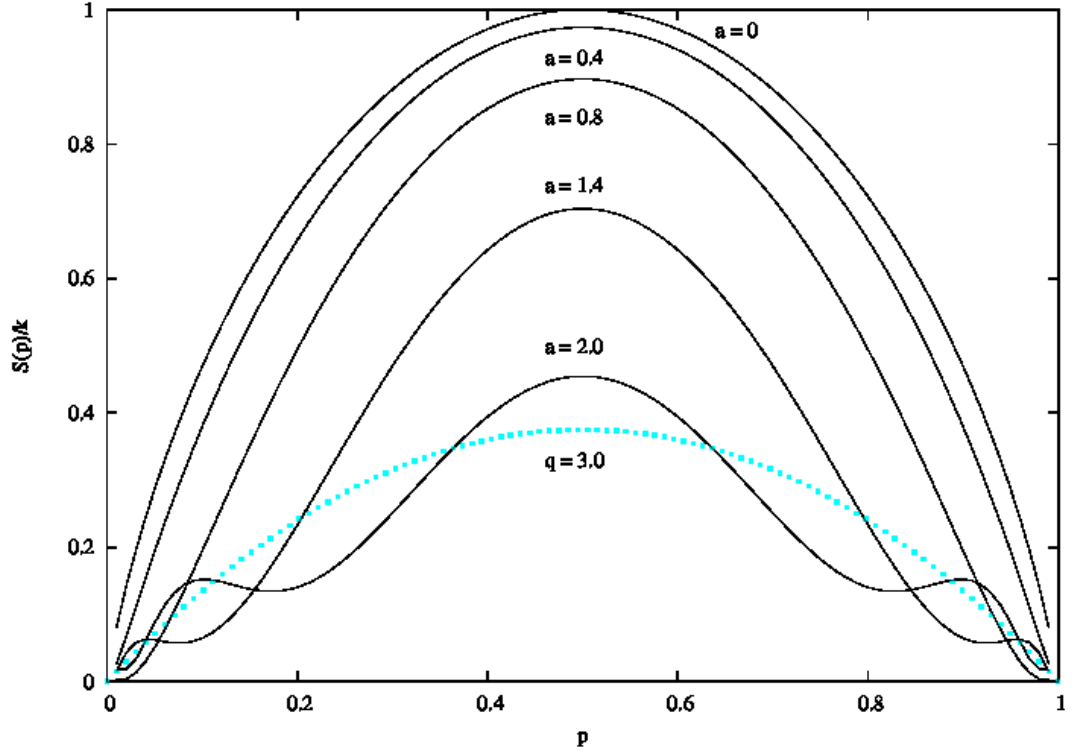}
\caption{Plot of $S$ in eq.(\ref{eq:logperiod}) for $N=2$ for different values of $|a|<\pi$. The familiar 
Boltzmann-Gibbs-Shannon, and  R\'{e}nyi forms are recovered in the limit $a \rightarrow 0$. Dotted lines corresponds 
to the Tsallis entropy function with $q \equiv 1+|a| = 3$.}
\label{fig:Fig1}
\vspace{1cm}
\end{figure}

As an illustrative example we show in fig.\ref{fig:Fig1} the binary entropy function ($N = 2$) using eq.(\ref{eq:logperiod}). 
For decreasing $a$, it attains its maximum value at $p = 1/2$. At intermediate values of $a$, there are measures of entropy
near to one relative to $S^{G}$. Log-periodicity in $S$ is manifested by increasing the value of our coefficient $a$.  
To this end, we mention that a rather similar 'anomalous' oscillatory behavior with maximum and minimum values for $S$ 
has been reported earlier in \cite{Bur70} for the vibrational entropy of small clusters of (40 to 80) atoms.
Furthermore, Tsallis has shown that concavity is also lost within a 2-parameter adaption of his non-extensive
entropy function $S_{q,\delta}$ ({\it c.f.}, eq.(\ref{eq:tsallisq})) \cite{Tsa13}.

Maximun and minimum peaks appear on $S$ when the derivative of our binary entropy function
\begin{equation}
dS(p)/dp = - (k/a) \{ \sin(a \log_{2} p) - \sin(a \log_{2} (1-p)) \}  \;\;\; ,
\end{equation}
becomes zero, with the logarithms in this formula taken to be base 2.
In the limit $a \rightarrow 0$ this relation tends to the negative of the {\it logit} function $(1/k) \log_{2} (p/(1-p))$, which corresponds to the 
logarithm of the odds in the Shannon information theory \cite{Sch96}.

Let us consider next the upper limit for the microstate normalized probabilities
\begin{equation}\label{eq:limitp}
0 < p^{|a|} = e^{\ln p^{|a|}} \approx 1 + |a|\ln p  \le 1 \le e  \;\;\; .
\end{equation}
It follows then that for positive values of $a<\pi$, $\sin(a\ln p) \approx \sin(p^{a} - 1)$ and $p^{a} - 1 \le 0$.
Therefore, we deduce from the discrete eq.(\ref{eq:logperiod}) and the first term in the above exponential series 
expansion that
\begin{equation}\label{eq:tsallisq}
S/k \approx - \sum_{i=1}^{N} p_{i} \sin(p^{a} - 1)/a \approx 
 \sum_{i=1}^{N} p_{i} (1 - p^{a})/a \approx \{ 1  - \sum_{i=1}^{N} p^{a+1} \} / a \;\;\; ,
\end{equation}
which corresponds to the non-additive Tsallis' $q$-entropy relation:
$\left (1  - \sum_{i=1}^{N} p^{q} \right)/(q-1)$, with a degree of non-extensivity $q - 1 \equiv a$.  
In fig.\ref{fig:Fig1} this limiting entropy function is plotted with dotted lines for $q = 3$.

From eq.(\ref{eq:limitp}) it also follows that $\ln p^{|a|} \le \ln e = 1$. Hence, 
$\sin^{2}((a/2)\ln p) \approx ((a/2)\ln p)^{2} \le (a/4)\ln p \approx (1/4) \sin (a\ln p)$. Therefore the non-additivity 
rule of eq.(\ref{eq:nonadd}) after some little algebra satisfies
\begin{eqnarray}\label{eq:nonaddtsallis}
S^{A+B} & \approx  &   
   S^{A} + S^{B} - (1/2) \left\{ S^{A} \cdot \sum_{j=1}^{N_{B}} p_{j}^{B} \sin (a\ln (p_{j}^{B}) )     + \; S^{B}\cdot \sum_{i=1}^{N_{A}} p_{i}^{A} \sin (a\ln (p_{i}^{A}) )  \right\}  \;\;\; , \nonumber \\
 & \approx  &   S^{A} + S^{B} - (a/k) \; S^{A} \cdot S^{B} \;\;\; ,
\end{eqnarray}
which is equivalent to Tsallis non-additivity \cite{Tsa88,Tsa09}:
$S_{q}^{A+B} = S_{q}^{A} + S_{q}^{B} + (1-q)\; S_{q}^{A} \cdot S_{q}^{B}/k$.
In this way we have extended the work of Tsallis \cite{Tsa88}. The first order term in the 
expantion of our measure of entropy complexity is in accord with the qualitative properties of the non-extensive 
$q$-formula, including the entropy of combined systems.  
The parametric family of Tsallis entropy measures preserve the additivity for independent events with $q = 1$.
It preserves most relevant features of the Boltzmann-Gibbs statistical mechanics formalism and is compatible with 
Shannon's Informaton theory foundations.

On the other hand, from the relation for the continous log-periodic entropy in eq.(\ref{eq:differential0}), it follows that 
for small $a$ and invariance $\sigma$, the expected value of $sin(a \ln p)$
--and of the continous entropy $H$, converges to the well-know 
$\ln \sqrt{2\pi\sigma^{2}} + 1/2 = \ln \sqrt{2\pi e\sigma^{2}}$ value for $H^{G}$ (and of the equivalent 
diferential Shannon entropy). This is the expected value to happen on average within the Gaussian distribution
independently of the mean parameter $\mu$.

Let us also consider Jensen's inequality for any convex function: 
$f\left( \sum_{i=1}^{N} \mu_{i} \cdot x_{i} \right) \le \sum_{i=1}^{N} \mu_{i} \cdot f(x_{i})$, where 
 $0 \le \mu_{i} \le 1$.  Hence, in the limit $\sin(a\ln x) \approx a\ln x$, we have that eq.(\ref{eq:logperiod}) 
satisfies the inequality
\begin{equation}\label{eq:renyis}
S/k \approx  - (1/a)  \sum_{i=1}^{N} p_{i} \ln p_{i}^{a}  \ge - (1/a) \ln \left( \sum_{i=1}^{N} p_{i} \cdot p_{i}^{a} \right)
  = \ln \left( \sum_{i=1}^{N} p_{i}^{\alpha} \right) /(1-\alpha) \equiv S^{R}\;\;\; ,
\end{equation}
where $S^{R}$ denotes R\'{e}nyi entropies and $\alpha -1 \equiv a$. The $\log$ function is external to the sum compatible 
with the axioms of probability. R\'{e}nyi entropies also contain Shannon as a special case \cite{Ren88}.


\section{Conclusions}

We have studied in this paper some properties of the log-periodic entropy, a quantity introduced by Sharma-Taneja in 1977. This quantity has received 
little attention in the literature. This is not due to the fact that this alternative quantity does not have any useful applications, but it is because 
only recently log-periodic functions have been of interest as applied to a variety of complex phenomena with variable mass.  We have remarked that 
the concept of entropy  does not have a unique definition but many.  Of these, the Shannon entropy has proved to be very successful for systems 
with constant mass.  We have considered systems with linearly increasing mass and have explained how "oscillating entropy" in such systems may appear. 

The concept of entropy plays important roles in diverse fields ranging from classical thermodynamics, quantum mechanics, cosmology
and information technology. However, it does not seem to have a unique definition. We have been interested here in its
possible log-periodic relation to microstates and their probabilistic approach, and we have made a connection
with the non-extensive Tsallis, R\'{e}nyi, Boltzmann-Gibbs and Shannon entropies as special limiting cases.
The presence of oscillations in the proposed entropy function for complex systems could be a manifestation 
of the presence of multiple sources to affect $p_{i}$.
This log-periodicity in our entropy, the concavity lost shown in fig.\ref{fig:Fig1}, and the non-additivity given in eq.(\ref{eq:nonadd})
are manifested by increasing the value of the coefficient $a$ in eq.(\ref{eq:logperiod}), which sets the variations with respect to the 
behavior of the monotonic Gibbs entropy function as in eq.(\ref{eq;agibbs}). 

There are two ways for the non-zero probabilities to change within an isolated system, both of them being irreversible in nature
\cite{Guj13}. One possible cause could be due to chemical reactions among microstates that brings about temporal
equilibration in the system. The other cause of probability changes is of quantum nature.  In this regard, we point out
that the evolution of the one dimensional classical and quantum oscillator systems
with fluctuating mass and spring constant
$H = \hat p^{2}/2m(t) + k(t)\hat q^{2}/2$
leads to temporal log-periodic oscillations \cite{Can09,Oze10}.

In analogy with a simple spring system gaining momentum from the surroundings as described by eq.(\ref{eq:force}), the expected values and the
evolution of entropy with time for our microscopic system could in principle be modeled assuming a linear-time dependent
behavior for the associated probabilities in order to understand how the entropy tends to its final value. Our model
could provide a rationale into the posssible log-periodicity on entropy and its concavity lost.  It may also help to figure out the nature of
entropy from a new perspective. Systems having log-periodic character are far from the equilibrium and unstable, and the total
mechanical energy of such oscillators reaches a minimum value in a sufficient amount of time \cite{Oze10}.
Furthermore, since entropy and heat capacity are so intimately related \cite{Val98,Soa03,Tsa12}, via
$C_{q}\partial T = T\partial S_{q}$, the complexity of systems as addressed by our oscillatory dynamical description
of the entropy could thus be further investigated at a fundamental level. Clearly interesting is to
apply this new log-periodic equation for the entropy in some specific cases. Work along these lines is in progress.



\end{document}